\documentstyle{mn}
\input{psfig.tex}
\newcommand{\phd}{\phantom{0}}
\newcommand{\php}{\phantom{.}}
\newcommand{\phm}{\phantom{-}}
\newcommand{\ts}{\thinspace}
\title[Lens Model for B1933+503]
   {Modeling the ten-image lensed system B1933+503}
\author[S.~Nair]
{S.~Nair$^{1,2}$
\thanks{Present address: Astrophysics Group, Raman Research
Institute, C. V. Raman Avenue, Bangalore 560080, India.}  \\
\phd$^{1}$Kapteyn Astronomical Institute, P.O. Box 800, 
9700 AV Groningen, The Netherlands  \\  
\phd$^{2}$University of Manchester, NRAL Jodrell Bank, Macclesfield, Cheshire
SK11 9DL, United Kingdom \\ }

\date{March 7, 1998}

\begin{document}

%\label{firstpage}

\maketitle

\begin{abstract}

A gravitational lens model is presented for the newly discovered
10-image system B1933+503. The underlying object, revealed by
modeling, is a triple radio source on the scale of a couple of hundred
mas that is well-aligned along the line of sight with a foreground and
somewhat flattened lensing galaxy, whose orientation and location
match that of an observed galaxy, known to be at a redshift of
0.755. Uncertainties in the modeling are obtained by a Monte Carlo
exercise. Observational tests of the lens model are proposed, and the
time delays between various pairs of images are determined as the core
of the source is known to be significantly variable.  Future
observations of the lens hold the key to using B1933+503 to constrain
Hubble's Constant. Despite the absence of a source redshift, the
system's utility as a probe of the lens galaxy's structure is
unparalleled as it provides a surfeit of easily identifiable
constraints for modeling the system.

\end{abstract}


\vskip 5truecm
\newpage

\section{Introduction}

In a companion paper, Sykes et al. (1997) report the discovery of a
spectacular arcsecond-scale gravitationally lensed radio system,
B1933+503 (1934+504) in J2000 coordinates), which was found in the
course of the Cosmic Lens All-Sky Survey (CLASS). CLASS is a search
for radio sources that exhibit multiple components with flat radio
spectral indices ($\alpha < 0.5$, with flux density $S_{\nu} \sim
\nu^{- \alpha}$). Such systems are typically extragalactic, and result
from gravitational multiple imaging of a background source such as a
quasar or an AGN by a foreground galaxy lying along the line of sight
to it. The first gravitational lens to be identified on the basis of
its multiple flat-spectrum radio components was PKS1830$-$211, by Rao
\& Subrahmanyan (1988). A survey conducted from Jodrell Bank (Patnaik
et al. 1992), now known as the Jodrell VLA Astrometric Survey or JVAS,
employed this property as a highly successful filter for identifying
new lensed systems: B0218+357 (Patnaik et al. 1993), B1422+231
(Patnaik et al. 1992), B1030+074 (King \& Browne 1996 and Xanthopoulos
et al. 1997) and B1938+666 (King et al. 1997a, King et al. 1997b), in
addition to the promising candidate lensed system B2114+022 (Augusto
et al. 1996). The ongoing CLASS project follows on from and contains
JVAS, and together these have discovered more than 13 systems so far
(inclusive of candidates yet to be confirmed) in a survey of nearly
8000 sources. Descriptions of this survey are to be found in Myers et
al.(1995), Jackson et al. (1995, 1997b) and Browne et
al. (1997). Newly established finds include B1608+656 (Myers et al.
1995, Fassnacht et al. 1996), B1600+434 (Jackson et al. 1995, Koopmans
et al. 1997a), B0712+472 (Jackson et al. 1997a), B1933+503 (Sykes et
al. 1997) and B1130+382 (Koopmans et al. 1997b).

B1933+503 bears a superficial resemblance to an Einstein Ring in its
morphology, but is composed of at least 10 discrete components, as
reported by Sykes et al. (1997).  It has been investigated over the
range of radio frequencies $1.7-15$~GHz (VLA 5, 8.4 \& 15~GHz, MERLIN
1.7 \& 5~GHz and VLBA 1.7 \& 5~GHz); only the MERLIN 1.7~GHz
observations appear to have the combination of dynamic range and
resolution required to pick up all 10 components. However, Sykes et
al. (1997) provide sufficient observational evidence to note that
(a)~four of the components have similar, complex radio spectra peaking
around 5 GHz and are compact, being a few mas in size, (b)~of the
remaining six, four have steep radio spectra over the range of
frequencies observed, but appear to be fairly compact (also on the
scale of a few mas). The remaining two components probably also have
steep radio spectra. Sykes et al. (1997) also infer that the
components which have complex radio spectra appear to be variable by
as much as 33\% at 15 GHz over the timescale of a couple of
months. HST observations with WFPC2 succeed in picking up what appears
to be a flattened galaxy located near the centre of the ring of radio
components. This object has an $I$ magnitude of $20.6$ at 810~nm and
is elongated at a position angle of $-40 \pm 5$ degrees, with an axial
ratio {\it (b/a)} in the plane of the sky of $0.45 -0.55$.  At 540~nm,
there is no sign of the galaxy down to V magnitude 22.5. An optical
spectrum taken with the Keck telescope using the Low Resolution
Imaging Spectrometer (LRIS) shows the presence of absorption and
emission corresponding to a redshift of 0.755, which probably
corresponds to the galaxy rather than the source since no optical
counterparts of the images have been detected as yet (Sykes et
al. 1997).

\section{B1933+503: the lens interpretation}

A galaxy acting as a lens typically produces five, three or a single
image(s) of a background source, with increasing degree of
misalignment of the source in the plane of the sky from the line of
sight to the lens (see, e.g., Schneider, Ehlers \& Falco 1992 for a
review of the properties of gravitational lensing by galaxies). When
five or three images are formed, one occurs very near the centre of
the galaxy and suffers a high degree of demagnification, so
observationally there would appear to be either four images (a quad)
or two images (a double) of the source. Most radio sources discovered
so far that are lensed by galaxies and exhibit multiple compact
flat-spectrum features can be clearly classified as either quad or
double systems. B1938+666 (King et al. 1997) has two components in the
background source, one of which is quadruply imaged while the other is
a double. With its 10 features, B1933+503 is thus simplest understood
as a triple radio source, the individual components of which have been
multiply imaged into a quad, a quad and a double.

The MERLIN 1.7 GHz map of Sykes et al. (1997) (at top left in fig.1 of
that paper) is the starting point of the present modeling
exercise. Most of the components in this map can be grouped into one
or the other of two quad configurations. The first quad involves
features 2, 5 and 7. The elongated morphology of feature 2 may be
interpreted as a pair of images that are partially merged across the
tangential critical curve (see Fig.~1(a)). There appear to be two
strong peaks of flux density in feature 2 in the 1.7 GHz map of Sykes
et al. (1997), and these will be referred to as 2a (east) and 2b
(west) in the present work.  The second quad is formed by components
1, 3, 4 and 6.  Component 8, lying as it does within the circle of
images, cannot then be singly imaged; its counterpart image is to be
found in the faint feature 1a to the NE of component 1.

The two-image configuration formed by 1a and 8 is unusual in that 8 is
so bright by comparison to 1a, despite its being nearer to the lens
centre.  This provides an interesting constraint on the lens model,
since this relatively rare configuration can be obtained if image 8 is
derived from three images that have just barely merged with each other
into a single bright image.  The source that is imaged into 1a and 8
must lie just outside a cusp of the tangential caustic in the source
plane, but within the radial caustic. (This is illustrated in
Fig.~1(b), a result from the lens modeling described in Sects. 3 and
4).

Within this picture, the underlying source prior to imaging by the
lens would consist of a central radio core which shares the spectral
and morphological properties of images 1, 3, 4, and 6, and is thus
flat-spectrum and compact. The radio core is flanked by two steeper
spectrum radio features, the trio being found by actual modeling to
lie almost in a straight line, on the scale of a couple of hundred mas
(Fig.~1(b)).

In principle, an alternative classification of the images could group
2 (seen again as a pair of images), 7, and 4 together, and 1, 3, 6,
and 8 into a second quad. This would require 5 to be singly-imaged, as
also 1a, unless each of them is an element of a double with a
counterpart image within the circle of images which is too weak to be
mapped. We can discard this and related scenarios in view of the
support that the earlier classification receives from the radio
spectra in figure~2 of Sykes et al. 1997, the VLBA 5GHz map in
figure~1 (Sykes et al. 1997), and the pattern of variability discussed
in that paper. (Amusingly, the image classification described earlier
was worked out prior to any knowledge of the radio spectra or the VLBA
observations.)

\section{Lens modeling}

A modeling code has been developed that seeks a best-fit model for
B1933+503, following the methods of Kayser $\&$ Schramm (1988) and
Kochanek (1991). These employ a penalty function, which is minimised
over the parameter space of a parametrized lens mass model, to yield a
`best fit' lens.  Although the observed flattening of the lens
suggests that it could be a disk galaxy which would typically consist
of a disk, a spheroid and a halo, it is treated here as a single
component elliptical lens approximating the overall mass distribution.
The lensing galaxy is described by a non-singular isothermal
ellipsoidal mass profile (the `PIEMD' of Kassiola $\&$ Kovner
1993). The mass density distribution in the lens, $\rho$, follows the
form:
$$ \rho(m) = \rho_o / (1+(m/a)^2), \eqno(1)$$ 
where $a$ is the scale length of the mass distribution, $\rho_o$ is
the central mass density, and $m^2=x^2+y^2/(1-e^2)+z^2$ is the
semi-major axis of an ellipsoidal shell of constant mass density. The
lens is thus an oblate spheroid with axial ratio given by
$\sqrt{1-e^2}$, and is assumed to be viewed edge-on. (The $x-$ and
$y-$ axes are confined to the plane of the sky, the orientation being
set by modeling the lensed system; in the absence of concrete
information about the nature of the lens, this model approximates the
behaviour of either an elliptical or a spiral galaxy). The lens model
has six parameters, these being the coordinates of the lens centre in
the lens plane, a mass parameter describing the strength of the lens,
its scale length, ellipticity and orientation in the plane of the
sky. The mass parameter, $\sigma_m$, is related to the central density
and the scale length:\ts $\rho_o = 9 \sigma_m^2 / 4 \pi G a^2$.
  
The complex scattering function formalism of Bourassa, Kantowski $\&$
Norton (1973) and Bourassa $\&$ Kantowski (1975) is employed, which
permits analytical expressions to be obtained for the lensing action
of a spheroidal mass distribution. In this formalism, the lens
equation relating the source position in the plane of the sky, ${\bf
z}_s=x_s+{\bf i}y_s$, to that of its images, ${\bf z}_i=x+{\bf i}y $,
via a scattering function, ${\bf I}(x,y)$, is given by:
$$ {\bf z}_s={\bf z}_i - {4 G F_d \over c^2}{\bf I}^{\ast} \eqno(2).$$
${\bf I}^{\ast}$ denotes the complex conjugate of the scattering
function, which is proportional to $\sigma_m^2$. {\bf I}$^{\ast}$ is a
function of image position ${\bf z}_i$ as well as of the lens
parameters ${\bf p}_j$.  $G$ is the gravitational constant, and $F_d$
is the ratio of two angular diameter distances, that between lens and
source, $D_{ls}$, and that between observer and source, $D_s$ (since
the redshift of the source is unknown at present, $F_d$ is absorbed
into the lens parameter $\sigma_m^2$ and the image and source
positions are in angular units in the plane of the sky). The
scattering function for the density profile in Eqn.(1) is given by
expression (4.1.2) of Kassiola $\&$ Kovner (1993). 

\subsection{Constraints on the Modeling}\label{constr}

Details of the image positions, relative to an initial guess lens
centre, are listed in Table 1, in which the images are grouped
according to their membership of a particular image configuration.
Each multiple image configuration of N images provides 2(N-1)
positional constraints (eliminating the common source position between
pairs of images in a group; see Eqn.(2)).  Thus, the two quads and the
double image configuration of B1933+503 supply a maximum of 6+6+2=14
constraints. This surfeit of constraints over parameters permits the
luxury of ignoring the relatively uncertain positions of images 2a and
2b: these images straddle the tangential critical curve and each
should theoretically be significantly brighter than either 5 or 7,
which share the same source but are removed from any critical curves.
This source, being of steep radio spectrum, is presumably non-variable
(note the constancy of the flux ratio for images 7 and 5 in Table 1;
the observations have been taken at different epochs). However, from
table 2 of Sykes et al. (1997) it is apparent that in almost every
observation, the {\it combined} flux density of 2a and 2b is lower
than that of 5 or 7. One is led to conclude that the source of the
image set (2a, 2b, 5, 7) is extended and is only partially imaged in
the vicinity of the tangential critical curve. Thus both the positions
of 2a and 2b and their flux densities are undependable inputs to the
modeling. Nevertheless, the fact that the image parities must be
reversed between the positions of features 2a and 2b (see Table 2)
proves very useful in constraining the location of the tangential
critical curve during the modeling process, and this parity
information is usable.  Thus, only 2+6+2=10 {\it positional}
constraints are actually employed.

Ratios between the observed flux densities of pairs of images in a
particular configuration generally provide an effective set of
constraints on the modeling. However, if the source shows temporal
variations in flux density, such changes manifest at different times
in the various images owing to differential time delays in the arrival
time at the observer for light from them. In B1933+503, it is known
that the quad images (1, 3, 4, 6), arising from the flat-spectrum
core, are significantly variable as discussed in Section 1. Hence
constraints derived from their flux densities at a given epoch of
observation could be uncertain by as much as 40\% in the higher
frequency observations. Accordingly, these are not be used in the
present modeling exercise.  Of the quad (2a, 2b, 5, 7), flux density
ratios involving 2a and 2b are neglected, as described earlier, but
the ratio for image 7 to image 5 appears to be robust (see Table 1);
the source is of steep-spectrum and is not expected to be variable.
Similarly, the flux density ratio of image 8 to 1a should be
non-variable, but in practice this is affected particularly by the
uncertainty in 1a, on account of its faintness.  No flux density ratio
constraints are actually employed in the modeling, but a good model
may reasonably be expected to reproduce the ratios 7/5 and 8/1a.
 
In the absence of reliable image flux density ratios, the image {\it
parities} provide important constraints on the modeling. The parity
$(-1)^n$ for an image in a given configuration (quad or double)
represents how the source maps onto that image, $n$ being the number
of reflections that it experiences about a set of coordinate axes
centred on the image (for a single lens plane, no rotations occur).
The theoretically expected image parities can be easily worked out by
inspection for standard elliptical lenses (see Blandford $\&$ Narayan
1986, for example), and are listed in Table 2 for B1933+503. This set
of constraints essentially determines the lens orientation in the
plane of the sky.

\subsection{The Penalty Function}

For the desired lens model, the positions of those images belonging to
the same quad or double configuration must map back through the lens,
via the lens equation, to their shared source position. The penalty
function to be minimised is obtained in the following manner~: In
general, having guessed an initial set of lens parameters, image
positions corresponding to the same quad or double do not map back via
Eqn.(2) to the same source position. A summed squared mismatch between
the various source positions obtained by backprojecting the observed
image positions, taken pairwise and divided by the number of images
involved, forms the first part of the penalty function; this should
ideally be as small as possible for each set of quads and the
double. In the literature, it has been popular to multiply the errors
in source recovery (departures from the average recovered source
position) by the model magnifications of each image, converting these
errors to values that can be compared directly with the observational
errors (e.g. Kochanek 1991). The minimization of the error function is
then carried out in the image plane. In the present work, we have
avoided using the model magnifications to bias the penalty
function. This is with good reason: trials in which it was attempted
to minimize the mismatch in the image plane after multiplying the
source plane mismatches with the local image magnifications tended to
locate models with enormously high image magnifications ($\sim 10^4$
and greater). In fact, given the freedom that a mass model with a
finite scale length permits, the program tended to place the
tangential critical curve as close to the images as it could (by
centering the lens with respect to the images and making it as round
as possible).  Trying to minimize the {\it source plane} mismatch
appears to be free of this bias and the results accord with trials
with synthetic data from (known) lens models, obtained using a
root-finding algorithm to solve the lens equation for the image
positions corresponding to a given source.

The second part of the penalty function incorporates the image parity
constraints. Only those lens models are accepted which exactly match
the predicted pattern of image parities. This ensures that unphysical
configurations are not selected by the program. Accordingly, the
penalty associated with a parity mismatch is very high. The penalty
function reads as:
$$ \phi =  \sum_{N_k=1}^{3} \bigg ( {1 \over N_k} \sum_{i<j}^{N_k} 
{ \mid {\bf z}_{s,i} -{\bf z}_{s,j} \mid }^2/{\sigma}^2 $$
$$ + \sum_{m=1}^{N_k} 10^4 \Gamma(P_mP_{em}) \bigg ), \eqno(3) $$
where $N_k$ denotes the number of images in the image group (quad or
double), $\sigma$ is the tolerance in recovering the source position,
and $P_m$ and $P_{em}$ are the model and expected values of parity for
image $n$ in a given group. The parities take values of either $+1$ or
$-1$.  $\Gamma(P_mP_{em})$ is a function with value zero for
$P_mP_{em}>0$, else it is unity. The large penalty associated with
this part of the function $\phi$ ensures that parity match is obtained
almost at the outset of the minimization process.

Minimization of the penalty function is achieved using a program,
`simann.f', written by Goffe (based on Goffe et al. 1994, and
available from the NETLIB public access site on the Internet at the
location http://www.netlib.org/opt/simann.f).  This program employs
the method of simulated annealing for optimization. It is found to be
particularly useful in accomodating the sudden discontinuities that
arise in the penalty function, accompanying changes in the model
values of image parity as the six-dimensional parameter space is
searched. Although computationally more expensive than the more
popular simplex method\footnote[4]{Descriptions of both simulated
annealing and simplex methods in optimization routines are to be found
in Press et al. (1992)} based optimization algorithms, it is
significantly less prone to getting trapped in local optima.

\section{Results and a discussion of the lens model}\label{result}

The source positions, as found for the best-fit ($\phi=9.83$ with a
target $\sigma$ of 5 mas) lens model are displayed in Table 2, and are
plotted in Figure 1(b). Note particularly the success of the model in
recovering the lobe double's source to lie just outside the cusp of
the tangential critical curve, as mentioned in Section 2. The model
image magnifications are also given in Table 2.  The parameters of the
best-fit model are given below; the errors quoted in brackets
alongside are 90\% confidence intervals from a Monte Carlo exercise.
For this, the image positions have been subjected to random
deviations, the magnitudes of which follow a normal probability
distribution with a standard deviation of 5 mas from their nominal
values. In the present case, 10000 such image configurations were
sampled and put through the modeling process described above, with all
six lens parameters free to vary. This took a total of 65 hours cpu
time on a Sparc Ultra 1 machine. While 10$^4$ trials may seem a sparse
sampling of the possible configuration space of random deviations in
all the image positions put together, in practice the estimated
confidence intervals alter only marginally between experiments with
$10^4$ trials and half that number (though a significant change occurs
between $500$ and $5000$ trials).

The best-fit model obtained by the methods described in the previous
sections yields lens parameters as below (with 90\% confidence
intervals quoted alongside):

\begin{tabbing}
 Scale Length $a$: $113\;{\rm mas}\;(59,224)$mas \\
 Eccentricity $e$: $0.81\; (0.73,0.84)$ \\
 Mass Parameter $\sigma_m$: $79.8\; (78.2,84.1)$ km/s \\
 P.A. of Lens Major Axis: $-46.5^{\circ}\;(-47.0^{\circ},-46.1^{\circ} )$ \\
 Lens Centre (RA, Dec. wrt cpt. 4): $(423^{+9}_{-5}, 270^{+7}_{-5})$ mas \\
\end{tabbing}

Histograms of the distributions of the mismatch function $\phi$ and
the various lens parameters, as obtained from the Monte Carlo
exercise, are presented in Figure 2. The `shoulder' in the plot for
the mismatch function seen around the values of 12 to 13 is probably
related to a similar feature seen in the plot for the position angle
around $-46.4$ to $-46.3$ degrees, which is likely to be a consequence
of the image parity constraints (sudden discontinuities result when a
change of parameters places an image in a region of image space with
the wrong parity).

The circular velocity $V_{c}(R)$ in the equatorial plane of the oblate
spheroidal lens galaxy with mass density distribution as in Eqn.(1),
can be estimated as a function of radial coordinate $R$ from
expression (2-91) in Binney $\&$ Tremaine (1987).
$$ V^2_c(R) = 9 \sigma_{m}^2 {\sqrt{1-e^2} \over e}{D_s \over D_{ls}}
 \bigg \{ {\pi \over 2} - {\rm tan}^{-1}{\sqrt{1-e^2} \over e}
 \mbox{\qquad \qquad} $$
$$ \mbox{ \qquad \qquad}-{1 \over \sqrt{1+R^2 / (ae)^2}}\;{\rm
cos}^{-1} \bigg ( {\sqrt{1-e^2} \over \sqrt{1+(R/ a)^2}} \bigg ) \bigg
\} \eqno(4)$$ In the limit of $e \longrightarrow 0$, the right hand
side of Eqn.(4) reduces to $ 9 \sigma_{m}^2 (D_s / D_{ls}) \{ 1 -
(a/R){\rm tan}^{-1}(R/a) \}.$ Asymptotically, the right hand side of
Eqn.(4) is: $9 \sigma_m^2 ({\sqrt{1-e^2} / e})(D_s / D_{ls}) \{ \pi /2
- {\rm tan}^{-1} ( \sqrt{1-e^2} /e ) \}$.  From the above, it is seen
that the circular velocity at a radius of $1^{\prime \prime}$ from the
lens centre is $183 \sqrt{D_s/D_{ls}}$ km/s, which, as $R$ tends to
large values, achieves a maximum of $198 \sqrt{D_s/D_{ls}}$ km/s.
 
The eccentricity $e$ translates to a flattening or b/a ratio in the
lens plane of $0.59^{+0.09}_{-0.05}$.  This is marginally rounder than
the value of $0.45-0.55$ suggested by the light profile in the HST
I-band image, which is not surprising if the degree of flattening of
the non-luminous component of the lens galaxy is somewhat smaller than
that of its disk. It is of interest that the orientation of the
lensing galaxy (p.a.$-40\pm 5$ degrees) accords so well with that
required by our lens model. This strongly suggests that B1933+503 is
lensed in the main by an isolated disk system. From Table 2, a
comparison of the image flux density ratios with the model
magnification values shows that the model actually predicts quite
reasonably the flux density ratio of components 7 and 5, and with a
lesser degree of success, the somewhat less reliable flux density
ratio 8/1a. In the case of the core images, departures of the model
magnification ratios from the observed flux density ratios are
apparent; these are probably the result of rapid intrinsic variations
in the core flux density, as discussed earlier.

\section{Predicted time delays for the images}\label{timede}

Time delays between the four images of the core of the radio triple
source, which we believe to be variable, are calculated for the mass
density profile in Eqn.(1) using the formulae of Cooke $\&$ Kantowski
(1975).  The time delay, $\Delta \tau$, for the arrival at the
observer of light from an image relative to, say, the undeflected
source, is the sum of the time delay due to differences in geometric
path length between the deflected and undeflected rays, $\Delta
\tau_g$, and that due to the apparent slowing of light by the
gravitational potential of the lens, $\Delta \tau_p$.  With the mass
density profile in Eqn.(1),
$$ \Delta \tau_g = \aleph {1 \over 2c} \big ({4 G \over c^2} \big )^2
\mid {\bf I}({\bf z}_i) \mid^2$$
$$ \Delta \tau_p = \aleph 36{\sigma_m^2 a \sqrt{1-e^2} \over c^3} {\rm
Re} \big \{ \int_0^{b_i} {({\bf z}_i/a)(\sqrt{1+b^2}-1) \over b
\sqrt{({\bf z}_i/a)^2 -(be)^2}}db \big \}$$ 
In the above expression,
the quantity $b_i$ is the value $(x^2+y^2/(1-e^2))/a^2$, where $x+{\bf
i}y={\bf z}_i$, the location of image $i$ relative to the lens centre
in the image plane.  $\aleph = (1+z_l)(D_l / F_d)$, which combination
of angular diameter distances has dimensions L$^{1}$ and calls for
additional input: the lens and source redshifts and a cosmological
model to be adopted. The integral in the expression for $\Delta
\tau_p$ is evaluated numerically.  Table 3 summarizes the time delay
results; for the sake of completeness the delays for images arising
from the lobes of the source are also listed. These are of relevance
if monitoring reveals relative motion between the core and the knots,
as is seen in compact symmetric objects (see Sec. 7.1 ahead).

\section{Fits with subsets of constraints}\label{subcon}

It is interesting to ask how the modeling behaves when only one set of
images is used. As discussed in Section 3.1, the positions of the four
core images provide by themselves a total of 6 constraints, which is
exactly equal to the number of model parameters. Not surprisingly, an
excellent match is obtained ($\phi_{CORE}=0.421$), with the scale
length restricted to be less than $10^{-5}$ asec (effectively a 5
parameter lens).  The smallness of the mismatch function $\phi$
probably points to some degree of redundancy in the information
provided by images in the core quad; the pairs 1 \& 4 and 3 \& 6 are
fairly symmetrically disposed with respect to the lens major axis
(Table~4).  The image magnification ratios 3/1, 4/1 \& 6/1 are 0.95,
1.87 \& 0.57 respectively. These should be compared with the observed
values of 0.69, 2.61 \& 0.61 for these images, recalling that they are
known to be highly variable, and with the original model values of
1.13, 2.08 \& 0.86 respectively.  (The value of $\phi_{CORE}$ for the
core images alone in the original model is 3.9 and for all the images
together is $\phi_{ALL}=9.8$).  However, if the scale length is
restricted and modeling is undertaken of all the images together, the
fit is a bit worse: $\phi_{ALL}=15.9$ for all the images together, or
$\phi_{CORE}=10.3$ for the core images alone. Save for the scale
length in these various models, the other parameters are similar; the
lens centre shifts by between 10 to 20 mas between models, and the
model velocity dispersion and eccentricity change by about 4\%, which
is within the 90\% confidence limits on the original model. The
orientation angle is almost unaffected. VLBA and MERLIN observations
underway (Sykes et al. 1997) should provide better determined
constraints from which to build more accurate models in the future.

\section{Discussion and Conclusions}

\noindent {\it Is B1933+503 a lensed system?} Despite the absence of
optical information on the images and a redshift for the source, the
simplicity with which the lensing picture explains this strikingly
unusual ten component system indicates that it is not just a
collection of physically related radio features but is indeed a lensed
source. This conclusion is underlined by the fact that actual lens
modeling accounts so well for details of the observations.

\smallskip
\noindent{\it The Mass Profile of the Lens Galaxy:} The mass model
adopted (a non-singular isothermal ellipsoid) is only an approximation
to possibly more complex substructure within the lens, especially if
it is a disk system as is suggested by the high degree of flattening
in the optical image. Since three source components are lensed by the
same galaxy, it will provide a strong test of the usual simplifying
assumptions made when modeling lensed systems. Although a quad system
provides a surfeit of constraints (6 positional + 3 image flux ratio =
9) over parameters for a simple 6 parameter lens model such as the one
used in the present work, the images typically tend to form at roughly
the same radial distance from the centre of the lens and convey little
information on the radial profile or the scale length the mass
distribution of the galaxy. This is particularly true if the the scale
length is quite small compared with the scale of the image
splitting. A double image system manages to sample radially distinct
portions of the lens, but in general on its own does not provide
enough constraints (2 positional + 1 image flux ratio = 3 quantities)
for even the simplest elliptical lens models.  B1933+503 is remarkable
in having two quads {\it and} a double, and the fact that the best-fit
model calls for a non-singular lens model (scale length of 113 mas,
translating to a typical value of about 0.5 kpc at $z_l=0.755$, H$_{\rm o}=
100$ km/s/Mpc) is
therefore a matter of some significance.

\smallskip
\noindent{\it Predictions for high-resolution observations:} The
constraints used in the present model are independent of the image
flux density ratios, but modeling relies upon the image parities,
which are robust constraints. The image parities are usually weak
constraints in a quad or a double image system, but in the present
10-image configuration effectively restrict the location of the
tangential critical curve.  The model in this paper makes certain
predictions: (i)~Magnification ratios for the core images are given in
Table 2, as obtained from modeling. Monitoring variations in the flux
density these images will provide both the time delays (also predicted
by the model, see Table 3) and the actual lens magnification ratios.
(ii)~The source of the quad (2a, 2b, 5, 7), a knot with a steep radio
spectrum, should itself be multicomponent or extended. In the present
model, this source is only partially imaged in the vicinity of 2a and
2b. VLBA observations at 1.7 GHz have been carried out (Marlow et
al. 1997), and these should reveal with greater accuracy the
correspondence between the peaks in 2a and 2b and features in the
other images 5 and 7.

\subsection{The source --- a Compact Symmetric Object?}

The source of this remarkable ten-image system appears to be a compact
radio triple source, an almost linear set of knot, core and a second,
weaker knot. The stronger and weaker knots have separations of 194.5
and 120 mas from the core in the present model (see Table 2). This
corresponds to a physical extent of about 1.3 kpc (H$_{\rm o}=100$
km/s/Mpc) and thus the source could be a compact symmetric object
(Readhead et al.  1994). Observations over time will indicate, by
changes in the locations and/or image flux densities of the knot
images, whether they are moving relative to the central core, as in
the case of the radio galaxy 1946+708 (Taylor \& Vermeulen
1997). While this would no doubt complicate the modeling described
earlier, it would in itself be a very interesting phenomenon to
observe.  It will be necessary to `deconvolve' the effects due to
differences in light travel time to the observer for the different
images from those due to the intrinsic source motion. Each epoch of
observation will catch upto four {\it different} phases in the time
evolution of the source structure in each quad, due to the relative
time delays between the images.  In addition, there is the advantage
of spatial magnification due to lensing, which can be as high in
specific directions as factors of 10$-$20 for images such as 2a, 2b,
and 8.  The model in the present paper makes an interesting prediction
in the event that the brighter knot moves outward from the core (see
Figure 1): the peaks of images 2a and 2b should disappear after
initially brightening and then merging with each other. Image 5 will
also move outward with respect to image 4.  If the fainter knot moves
outward from the core, image 8 should initially become fainter (it
moves into a region of lower magnification); there will be an increase
in separation between images 1 and 1a.

\subsection{B1933+503 as a cosmological probe}

A well-established goal of gravitational lensing is to determine
values of parameters for cosmological models (Refsdal 1964). From the
expressions in Section \ref{timede}, the factor $\aleph$ is inversely
proportional to Hubble's Constant; thus, with a well-fit model for the
lens and measured values for the time delays between pairs of images,
it is possible to estimate Hubble's Constant within a specific
cosmological model. However, the quantity $\aleph$ involves the source
redshift, $z_s$, which in the case of B1933+503 is unknown. A study of
the lens galaxy could provide the key to using this system to estimate
Hubble's Constant. From Section \ref{timede}, it is evident that
information involving the source redshift is restricted to the term
$F_d$. If it proves possible to observationally determine the circular
velocity $V_c$ of the lens through redshifted HI or molecular line
studies, then from Equation (4) and the model parameters, the quantity
$F_d=D_s/D_{ls}$ is obtainable in principle. This is under the
assumption that lensing is due to the galaxy at $z_l=0.755$, with no
significant contribution from external shear or matter along the line
of sight.  In view of the degree to which the simple model presented
here accounts for B1933+503, this does not seem unreasonable. In
general, the presence of distributed matter along the line of sight
cannot be ruled out, so at least an upper bound can be put on the
value of Hubble's Constant.

The treatment in this paper assumes that the lens is viewed edge-on;
including an inclination along the line of sight is straightforward
with regard to the lens model (see Bourassa \& Kantowski 1975), and
should observations provide sufficient detail to permit the
inclination to be established, B1933+503 will be a rather promising
system for cosmological studies.

\section*{Acknowledgments}
The author gratefully acknowledges discussions with Ian Browne and
Chris Sykes, with Roger Blandford regarding the lens modeling, Dan
Marlow and Peter Wilkinson regarding the VLBA observations, with Neal
Jackson regarding the HST observations of the lens, and with Rene
Vermeulen regarding the nature of the source.  Bill Goffe kindly
provided details of his optimization code. Many discussions with Leon
Koopmans made the work both stimulating and a pleasure, and the author
is grateful to Ger de Bruyn for discussions about future observational
possibilities with B1933+503.  The Astronomical Institute `Anton
Pannekoek', University of Amsterdam and the Netherlands Foundation for
Research in Astronomy, Dwingeloo, are gratefully acknowledged for the
use of their computing facilities, as is the pulsar group at Jodrell
Bank, on whose computers the initial simulations were run. The author
acknowledges NWO grant number B-78-344 while in the Netherlands.

%Ackn. NWO grant number B-78-344

\begin{table*}
\begin{tabular}{cccccc}
\hline
Image & Rel. \ts (RA, Dec)&
          \multicolumn{4}{c}{Obs. flux ratio at Freq.\ts (GHz) } \\
          & (mas)   & 1.7 & 5  & 8.4 & 15 \\
\hline
2a        & $(\phm 171 , \phm 414) $ & \ldots   & & & \\
2b        & $(\phm \phd 51 , \phm 420)$  & \ldots   & & & \\
5         & $(-531   ,   -497)$      & 1.00     & 1.00  & 1.00   & 1.00 \\
7         & $(\phm 398,  -134)$     & 1.25   & 1.15   & 1.16   & 1.13  \\
\hline
1         & $(\phm 447 , \phm 495)$  & 1.00 & 1.00  & 1.00   & 1.00  \\
3         & $(-389 , \phm 158)$  & 0.69   & 0.84  & 0.79   & 0.61  \\
4         & $(-397 ,   -299)$      & 2.61   & 3.46  & 3.65   & 3.78  \\
6         & $(\phm 230  , -387)$      & 0.61   & 0.96  & 0.83   & 0.78  \\
\hline
1a        & $(\phm 545  , \phm 584)$  & 1.00  & & & \\
8         & $(-114   ,   -335)$      & 4.00  & & & \\
\hline
\end{tabular}
\caption{Inputs to the modeling: For each set of images, the positions
relative to an initial guess lens centre at (397,299) mas wrt image 4,
and the image flux ratios relative to that image of each set whose
value is 1.00, for various frequencies of observation. The model uses
only the 1.7 GHz data; the other observations are included in the
Table to indicate the (un)reliability of various image flux ratios as
constraints. Values for images 2a \& 2b are not listed as it is
apparent that these are mappings of only a fraction of the source of
images 5 \& 7 (see text).  In the higher frequency observations, the
resolution and/or dynamic range was inadequate to pick up 2a \& 2b as
separate components.  Images 1a \& 8 are not detected in the higher
frequency observations. The core images 1, 3, 4 \& 6 show signs of
variability ({\it cf.} discussion in Sykes et al. 1997). Most data are
derived from Sykes et al. (1997); positions for 2a \& 2b are from
Browne (private communication).}
\end{table*}

\begin{table*}
\begin{tabular}{cclcl}
\hline
 Image & Model & Flux Ratio& $P_m$ &
          Source \\
 Set & {\rm Magn.} & {\rm (obsvd)} &($P_{em}$) & {\rm Pos. (mas)}\\
\hline
 \multicolumn{2}{l}{Lobe quad:}  & &  &\\
   5 & \phd $2.58$   & $1.00$  & $+1$ &  $\delta RA:$  \\
     &        &    $1.00$             & $+1$               & $-105\pm5$ \\
   7 & \phd $3.05$   & $1.18 \php(1.03,1.62)$  & $-1$ &  $\delta Dec:$ \\
     &        & $1.25 \php (\pm0.04)$  &   $-1$             & $-66\pm1$ \\
   2a& $\ldots$ & $\ldots$        & $+1$ &    \\
     &        &                 &     $+1$           &              \\
   2b& $\ldots$ & $\ldots$         & $-1$ &    \\
     &        &                 &  $-1$              &              \\
\hline
 \multicolumn{2}{l}{Core quad:}  & &  & \\
   1 & \phd $2.89$   & $1.00$  & $+1$ & $\delta RA:$ \\
     &        &    $1.00$           & $+1$               &  $\phd  36\pm3$ \\
   3 & \phd $3.25$   & $1.13 \phd(1.02, 1.42)$  & $-1$ & $\delta Dec:$ \\
     &        & $0.69 \php (\pm0.14)$ &     $-1$           & $\phd 68\pm4$  \\
   4 & \phd $6.01$   & $2.08 \phd(1.98,2.52)$  & $+1$ &    \\
     &        & $2.61 \php (\pm0.31)$ &   $+1$             &        \\
   6 & \phd $2.48$   & $0.86 \phd(0.75,1.16)$ & $-1$ &    \\
     &        & $0.61 \php (\pm0.13)$  & $-1$        &              \\
\hline
\multicolumn{2}{l}{Lobe double:}& &  & \\
   1a& \phd $2.28$   & $1.00$  & $+1$ & $\delta RA:$ \\
     &        &  $1.00$          &  $+1$              & $\phd 123\pm5$  \\
   8 & $14.24$  & $6.24 \phd (4.62,10.22)$ & $-1$ & $\delta Dec:$  \\
     &        & $4.00 \php (\pm1.83)$  & $-1$    &  $\phd 151\pm4$ \\
\hline
\end{tabular}
\caption{Model Predictions: For the components grouped as discussed in
Section 3, the model magnifications, model flux density ratios (with
90\% confidence intervals determined from a Monte Carlo exercise
alongside) --- in the succeeding line, these are compared with
observed values taken from the 1.7 GHz map of Sykes et al. 1997 (with
errors as inferred from that paper); the model image parities $P_m$
(with the theoretically expected values $P_{em}$ on the following
line), and the model-recovered source positions $(\delta({\rm
RA}),\delta({\rm Dec}))$ for each image group, w.r.t the lens
centre. The source positions are mean values for all backprojected
images belonging to a particular group (i.e., quad or double).  Beside
these values are the standard deviations for each set, for the
best-fit model itself; ideally, these should be as small as possible.}
\end{table*}

\begin{table*}
\begin{tabular}{lcc}
\hline
Image & Time Delay & 90\% Confidence Interval \\
Set & \multicolumn{2}{c}{(in units $\aleph$ days/Gpc)} \\
\hline
\multicolumn{2}{l}{Lobe quad:}  & \\
{\bf 5} & $0.00$ & \ldots \\
{\bf 7} & $4.32$ & $(3.56,\phd 4.89)$ \\
{\bf 2a}& $3.66$ & $(3.14,\phd 4.05)$ \\
{\bf 2b}& $3.83$ & $(3.20,\phd 4.29)$ \\
\hline
\multicolumn{2}{l}{Core quad:}  & \\
{\bf 1} & $0.00$ & \ldots \\
{\bf 3} & $2.74$ & $(2.11, \phd 3.13)$ \\
{\bf 4} & $2.37$ & $(1.95, \phd 2.71)$ \\
{\bf 6} & $3.15$ & $(2.52, \phd 3.58)$ \\
\hline
\multicolumn{2}{l}{Lobe Double:} & \\
{\bf 1a}& $0.00$ & \ldots \\
{\bf 8} & $6.22$ & $(5.15, \phd 7.02)$ \\
\hline
\end{tabular}
\caption{Predicted time delays for the three image sets: The image listed
at top in each set is the reference image corresponding to the global
minimum in arrival times for light from the source (core or a lobe).
$\aleph= (1+z_l)(D_l / F_d)$, as discussed in Section 5 (see text);
a typical value is 3 Gpc).
The last column lists the 90\% confidence intervals for each value, from the
Monte Carlo exercise.}
\end{table*}

\begin{table*}
\begin{tabular}{lcc}
\hline
Image & $B$ value & $\mid \psi \mid$ \\
Set & (asec) & (degrees) \\
\hline
\multicolumn{2}{l}{Lobe Quad:} & \\
5 & 1.24 & 83 \\
7 & 0.46 & 27  \\
2a & 0.75  & 65 \\
2b & 0.66 & 51 \\
\hline
\multicolumn{2}{l}{Core Quad:} & \\
1 & 1.14 & 86 \\
3 & 0.50 & 19 \\
4 & 0.80 & 75 \\
6 & 0.44 & 17 \\
\hline
\multicolumn{2}{l}{Lobe Double:} & \\
1a & 0.80 & 3 \\
8 & 0.37 & 19 \\
\hline
\end{tabular}
\caption{Illustrating the extent of sampling of the lens plane by the
various image systems. $B$ is the elliptical radius of the given
image: in Cartesian coordinates, referred to the lens centre and with
the $x-$axis along the lens major axis, $B^2=x^2+y^2/(1 - e^2)$. $\mid
\psi \mid$, the norm of the angle each image makes with respect to the
lens major axis (when referred to the lens centre), is listed
alongside. Images of similar $\mid \psi \mid$ and $B$ values provide
similar constraints on the modeling if they belong to the same image
system.}
\end{table*}

\begin{figure*}
\caption{Lens model for B1933+503. (a)~: Image plane critical curves
for the lens model, on which observed image positions have been
superposed. The lens is centred at (0,0). Images landing very near a
critical curve suffer very high magnifications, tending to merge with
nearby images across such curves (2a with 2b, for example). Image
positions marked by open circles with central dots form the quad
arising from lensing of the source's core, open circles mark one
lobe's quad images, and the small filled squares denote the other
lobe's double images.  (b)~: The corresponding source plane caustics
for the model. The positions of the three source components are marked
with symbols corresponding to their images, a circle with a dot for
the core of the source, an open circle for the stronger lobe and a
small filled square for the weaker one.  The cross superposed on the
open circle is the source location when the positions of images 2a and
2b are included (see Section 3.1); the slight offset with
respect to the original source position, obtained from images 5 and 7
alone, arises because images 2a and 2b represent only a fraction of
the source that gets mapped onto 5 and 7.}

\psfig{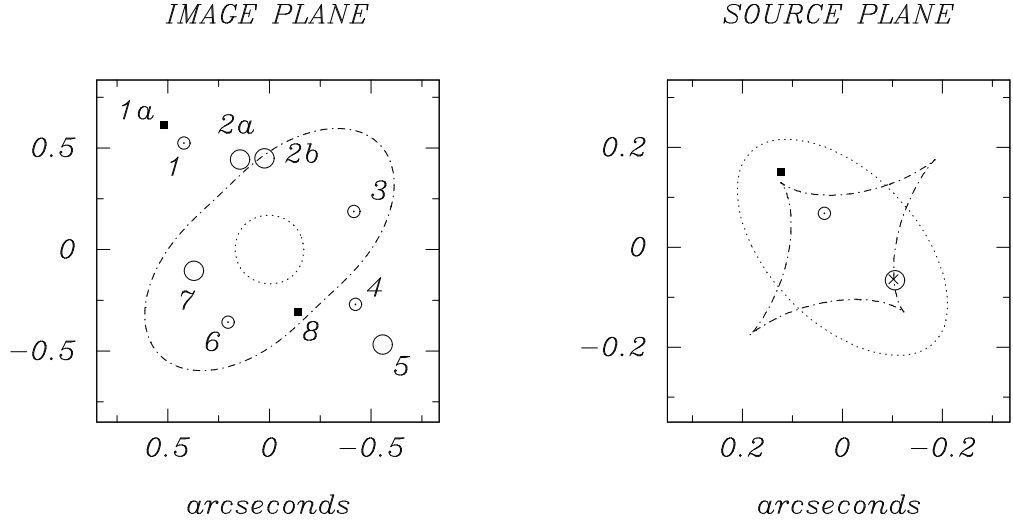}

\end{figure*}

\begin{figure*}
\caption{Results from the Monte Carlo exercise (Section 4).
The plots are histograms of binned values of the six
lens parameters, and of the mismatch function $\phi$.  The plot for
the mismatch function shows the buildup of the distribution as the
number of experiments, N$_{\rm exp}$, is increased from $10^3$ to
$10^4$.  Note that the lens parameters are varied {\it simultaneously}
in this set of experiments, which results in a significantly wider
spread in recovered values than is currently recognised in the
literature. A spread of the order of 30 \% is apparent in the
predicted time delay values as well (Table 3).}
\psfig{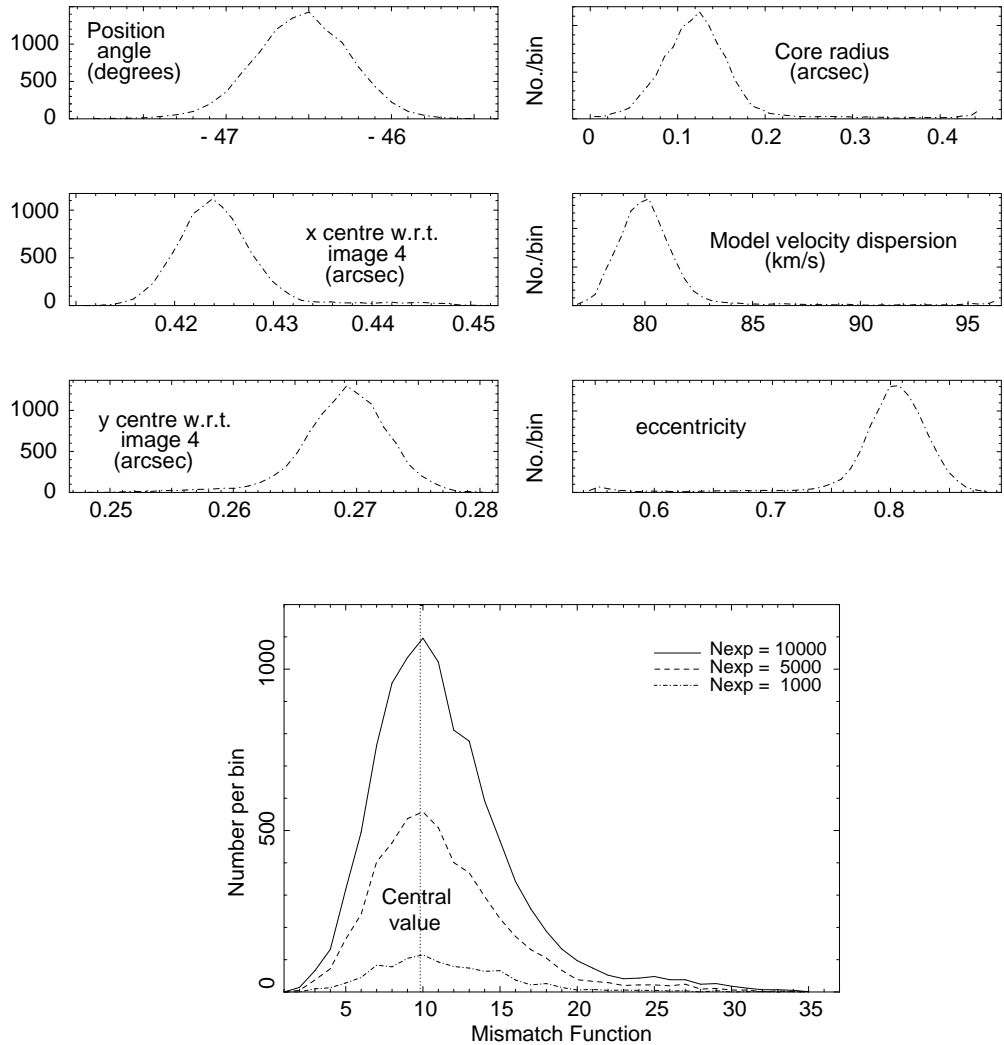}
\end{figure*}

\bsp

\begin{thebibliography}{}
\bibitem{}
Augusto, P. et al., 1996, in Astrophysical Applications of
Gravitational Lensing, eds. Kochanek, C.S \& Hewitt, J.~N.
(Dordrecht: Kluwer)
\bibitem{}
Binney, J. \& Tremaine, S. 1987, Galactic Dynamics, Princeton University 
Press, Princeton, N.J., USA
%\bibitem{}
%Blandford, R.D. \& Kochanek, C. 1987, ApJ, 321, 658
%\bibitem{}
%Blandford, R.D. \& Kovner, I. 1988, Phys Rev A, 38, 4028
\bibitem{}
Blandford, R.D. \& Narayan, R. 1986, ApJ, 310, 568
%\bibitem{}
%Grundahl, F., Hjorth, J., 1995, MNRAS, 275, L67
\bibitem{}
Bourassa, R.R., Kantowski, R. \& Norton, T.D., 1973, ApJ, 185, 747 
\bibitem{}
Bourassa, R.R. \& Kantowski, R., 1975, ApJ, 195, 13
\bibitem{}
Browne, I.W.A. et al., 1997, in `Observational Cosmology with the new radio
surveys', eds. Bremer, M., Jackson, N. \& Perez-Fournon, I. (Kluwer)
\bibitem{}
Cooke, J.H. \& Kantowski, R., 1975, ApJ, 195, L11
\bibitem{}
Fassnacht, C.D., Womble, D.S., Neugebauer, G., Browne,
I.W.A., Readhead, A.C.S, Matthews, K \& Pearson, T.J,
1996, ApJ, 467, 73
\bibitem{}
Goffe, W.L., Ferrier, G.D. \& Rogers, J. 1994, J. Econometrics, 60, 65
\bibitem{}
Jackson, N., et al., 1995, MNRAS, 274, L25
\bibitem{}
Jackson, N., et al., 1997a, MNRAS, in press
\bibitem{}
Jackson, N., Nair, S. \& Browne, I.W.A., 1997b, in `Observational 
Cosmology with the
New Radio Surveys', eds. Bremer, M., Jackson, N. \& Perez-Fournon, I. (Kluwer)
\bibitem{}
Kassiola, A. \& Kovner, I. 1993, ApJ, 417, 450
\bibitem{}
Kayser, R. \& Schramm, T. 1988, A\&A, 191, 39
\bibitem{}
King, L.J. \& Browne, I.W.A.  1996, MNRAS, 282, 67
\bibitem{}
King, L.J., Browne, I.W.A., Muxlow, T.W.B.,
Narasimha, D., Patnaik, A. R., Porcas, R.W.
\& Wilkinson, P.N.,  1997a, MNRAS, 289, 450
\bibitem{}
King, L.J., Jackson, N.J., Blandford, R.D., Bremer, M.N., Browne,
I.W.A., de Bruyn, A.G., Fassnacht, C., Koopmans, L., Marlow, D.,
Nair, S. \& Wilkinson, P.N., 1997b, submitted
\bibitem{}
Kochanek, C.S. 1991, ApJ, 373, 354 
\bibitem{} 
Koopmans, L., de Bruyn, A.G. \& Jackson, N., 1997a, MNRAS, submitted
\bibitem{} 
Koopmans, L. et al., 1997b, in preparation
\bibitem{} 
Marlow, D. et al., 1997, in preparation
\bibitem{} 
Myers, S., et al., 1995, ApJ, 447, L5
\bibitem{}
Patnaik, A.R., Browne, I.W.A., Walsh, D.,
Chaffee, F.H. \& Foltz, C.B., 1992, MNRAS, 259, 1P
\bibitem{}
Patnaik, A. R., Browne, I.W.A., King, L.J.,
Muxlow, T.W.B., Walsh, D. \& Wilkinson, P.N.,  1993, MNRAS, 261, 435
\bibitem{}
Press, W.H., Teukolsky, S.A., Vetterling, W.T. \& Flannery, B.P. 1992,
{\it Numerical Recipes in Fortran: The Art of Scientific Computing}, 
Cambridge Univ.
Press, Cambridge, Mass., USA
\bibitem{}
Rao, A. Pramesh \& Subrahmanyan, R., 1988, MNRAS, 231, 229
\bibitem{}
Readhead, A.C.S., Taylor, G.B., Xu, W., Pearson, T.J., Wilkinson, P.N.
\& Polatidis, A.G., 1994, ApJ, 460, 612
\bibitem{}
Refsdal, S., 1964, MNRAS, 128, 307
\bibitem{}
Schneider, P., Ehlers, J. \& Falco, E.E., 1992, Gravitational Lenses
(New York: Springer)
\bibitem{}
Sykes, C.M. et al., 1997, companion paper, astro-ph/9710358
\bibitem{}
Taylor, G.B. \& Vermeulen, R.C., 1997, ApJ, 485L
\bibitem{}
Xanthopoulos, E. et al., 1997, in preparation
\end{thebibliography}
\end{document}